# Metastable ferroelectricity in optically strained SrTiO$_3$

T. F. Nova[1,2,*], A. S. Disa[1], M. Fechner[1], A. Cavalleri[1,2,3,*]

[1]Max Planck Institute for the Structure and Dynamics of Matter, 22761 Hamburg, Germany

[2]The Hamburg Centre for Ultrafast Imaging, 22761 Hamburg

[3]University of Oxford, Clarendon Laboratory, Oxford OX1 3PU, UK

*Corresponding authors

**Fluctuating orders in solids are generally considered high-temperature precursors of broken symmetry phases. However, in some cases these fluctuations persist to zero temperature and prevent the emergence of long-range order, as for example observed in quantum spin and dipolar liquids. SrTiO$_3$ is a quantum paraelectric in which dipolar fluctuations grow when the material is cooled, although a long-range ferroelectric order never sets in. We show that the nonlinear excitation of lattice vibrations with mid-infrared optical pulses can induce polar order in SrTiO$_3$ up to temperatures in excess of 290 K. This metastable phase, which persists for hours after the optical pump is interrupted, is evidenced by the appearance of a large second-order optical nonlinearity that is absent in equilibrium. Hardening of a low-frequency mode indicates that the polar order may be associated with a photo-induced ferroelectric phase transition. The spatial distribution of the optically induced polar domains suggests that a new type of photo-flexoelectric coupling triggers this effect.**

$SrTiO_3$ is paraelectric and centrosymmetric at all temperatures. When cooled, it displays many anomalies that suggest its proximity to a ferroelectric phase, including a large rise in the dielectric function[1] and softening of a polar mode[2]. This behavior is often referred to as incipient ferroelectricity, with quantum fluctuations of the ionic positions preventing long-range ordering[3,4]. However, the proximity to a ferroelectric phase is underscored by the ease with which $SrTiO_3$ can be made ferroelectric, for example by Ca-substitution (Sr → Ca; $T_c$ = 37 K)[5] or by isotope substitution ($^{16}O$ → $^{18}O$; $T_c$ = 25 K)[6]. Strain, as shown in the phase diagram of Fig. 1, has proven most effective in controlling the transition in $SrTiO_3$[7], with reported ferroelectricity up to room temperature[8].

In analogy with the application of strain by epitaxial constraints or static pressure (see Fig. 1), we used mid-infrared optical pulses to drive infrared-active vibrations to large amplitudes in order to deform the lattice of $SrTiO_3$ and dynamically drive long-range ordering. *A priori*, nonlinear phononics[9,10,11] may induce a polar or ferroelectric phase in many ways. For example, dynamical phonon softening[12,13] of the polar mode by cubic lattice nonlinearities[14] or the generation of transiently-induced strain (Supplementary S11) may all provide routes to creating a ferroelectric order absent at equilibrium.

The highest-frequency $A_{2u}$ vibrational mode of $SrTiO_3$ was resonantly excited (at T = 4 K) with femtosecond mid-infrared optical pulses tuned to 15-μm wavelength (83 meV photon energy), derived from a 1 KHz repetition rate Ti-sapphire Laser, an optical parametric amplifier and a difference frequency mixing crystal (Supplementary S1). The electric field polarization was oriented along the [001] axis of a (110)-oriented crystal (Fig. 2A).

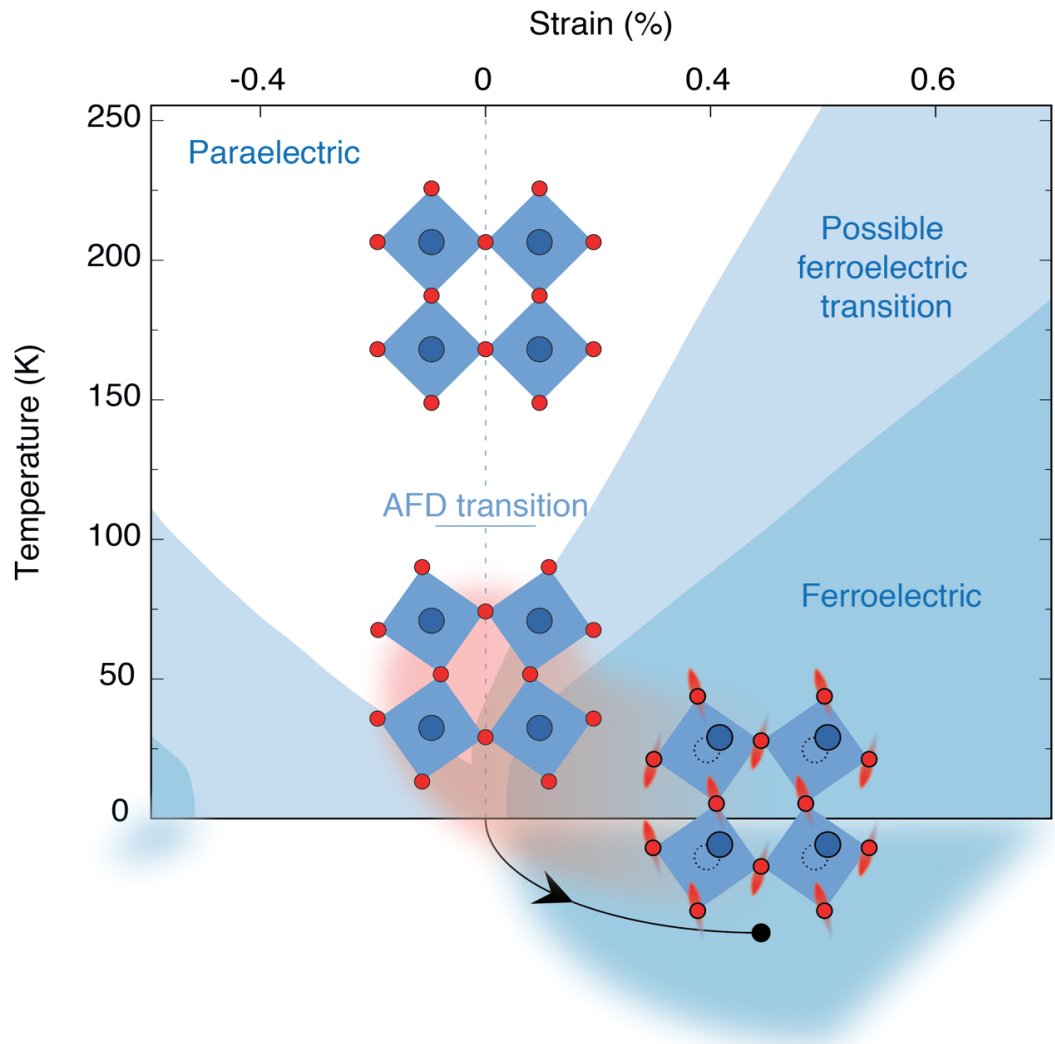

**Figure 1 | Dynamical strain in SrTiO₃.** Bulk, unstrained SrTiO₃ is paraelectric at any finite temperature. At 105 K it undergoes an antiferrodistortive (AFD) transition from cubic to tetragonal, although it retains a centrosymmetric structure. However, small amounts of strain cause the material to undergo a ferroelectric transition. The Curie temperature increases with growing values of applied strain. The strain-phase diagram shown here represents a thermodynamic analysis of the phase transition for a single-domain state, adapted from ref. [15]. The shaded cartoon explores the possibility of dynamically establishing a ferroelectric phase through vibrational excitation.

The symmetry of SrTiO₃ was monitored via second harmonic generation[16] of a 2.2 μm-wavelength optical probe pulse, collinear and time-delayed with respect to the mid-infrared pump (Fig. 2A). As shown in Fig. 2B a *time-delay-independent* second harmonic signal, absent without the pump, indicated the appearance of a non-centrosymmetric phase after mid-infrared excitation.

The second harmonic signal was observed to accumulate with exposure to mid-infrared radiation and reach saturation after several minutes, with a maximum value determined by the pump fluence (30 mJ/cm$^2$ for the data of Fig 2B; Supplementary S5). As all other contributions to the second harmonic were negligible, including surface and quadrupole terms of the nonlinear susceptibility tensor, this observation was a reliable reporter of a photo-induced phase with broken inversion symmetry (Supplementary S2).

Structural symmetry information was obtained by continuously rotating the incoming 2.2-μm probe polarization and measuring the 1.1-μm second harmonic projected along the pseudo-cubic [1-10] and [001] crystallographic directions with a second polarizer (Fig. 2C). These angular dependences indicate the formation of a polarization along the [1-10] direction and are consistent with a (non-centrosymmetric) polar point group ($C_{2v}$, Supplementary S2).

The experiment was performed as a function of sample temperature and pump wavelength. The induced second harmonic signal was found to be maximum at low temperatures; however, a detectable effect could be induced even up to room temperature (Supplementary S3). In addition, the polar state was found to be most efficiently created when pumping resonantly with the $A_{2u}$ phonon, while for pump photons in the proximity of the 3.2 eV band gap of $SrTiO_3$, the effect disappeared (Fig. 2A inset and Supplementary S6).

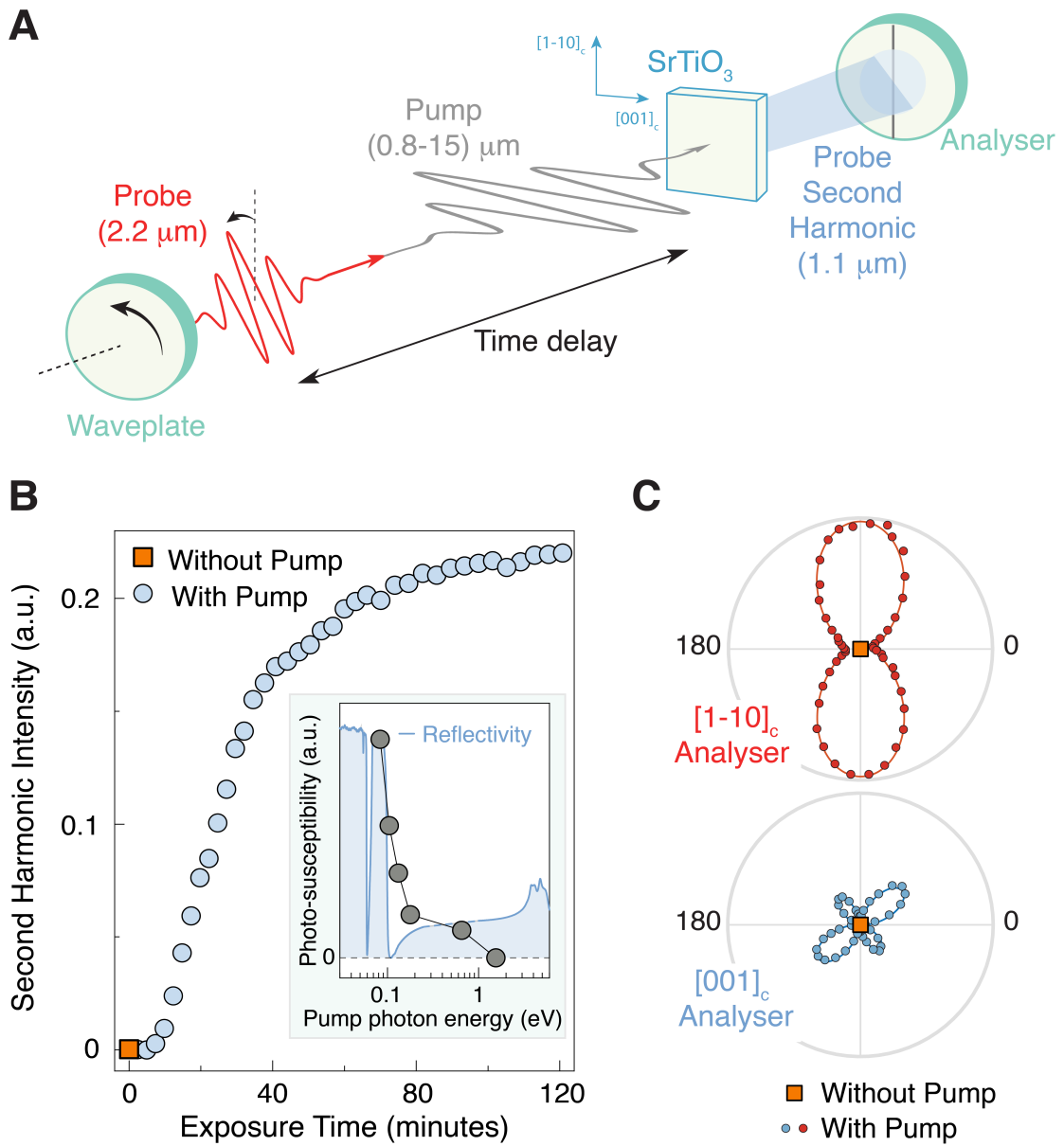

**Figure 2 | A photo-induced polar state. A**, Sketch of the experimental setup. A (110)-oriented SrTiO$_3$ sample is coherently excited with tunable wavelength pulses (0.8 – 15 μm, grey line). Time-delayed collinear probe pulses (2.2 μm wavelength) impinge on the sample, with a half-waveplate controlling their polarization. The generated second harmonic (1.1 μm) is detected in transmission geometry. When needed, an analyzer (i.e. a polarizer) can be used to isolate orthogonal polarization components of the second harmonic. **B**, Time-delay-independent total second harmonic intensity impinging on the detector (without analyzer) as a function of exposure time to 15 μm pump pulses. **Inset**, Pump wavelength dependence. The grey dots represent the photo-susceptibility of the effect, as defined in Supplementary S6. The data are compared to the static reflectivity of SrTiO$_3$ (blue line). The reflectivity low-energy data (< 0.2 eV) were measured by Fourier Transform Infrared Spectroscopy (FTIR) with a commercial spectrometer. The high-energy data (> 1 eV) were adapted from ref. [17]. **C**, Second harmonic intensity in the saturated state as a function of the probe polarization for two orthogonal analyzer configurations.

Interestingly, the photo-induced phase was found to be metastable when unperturbed, relaxing back to the non-polar equilibrium paraelectric phase only several hours after the pump was turned off (Fig. 3A). One can also revert the material to its paraelectric ground state by exposing it to above-band-gap photons[18] (> 3.2 eV, Fig. 3B) or by thermal cycling.

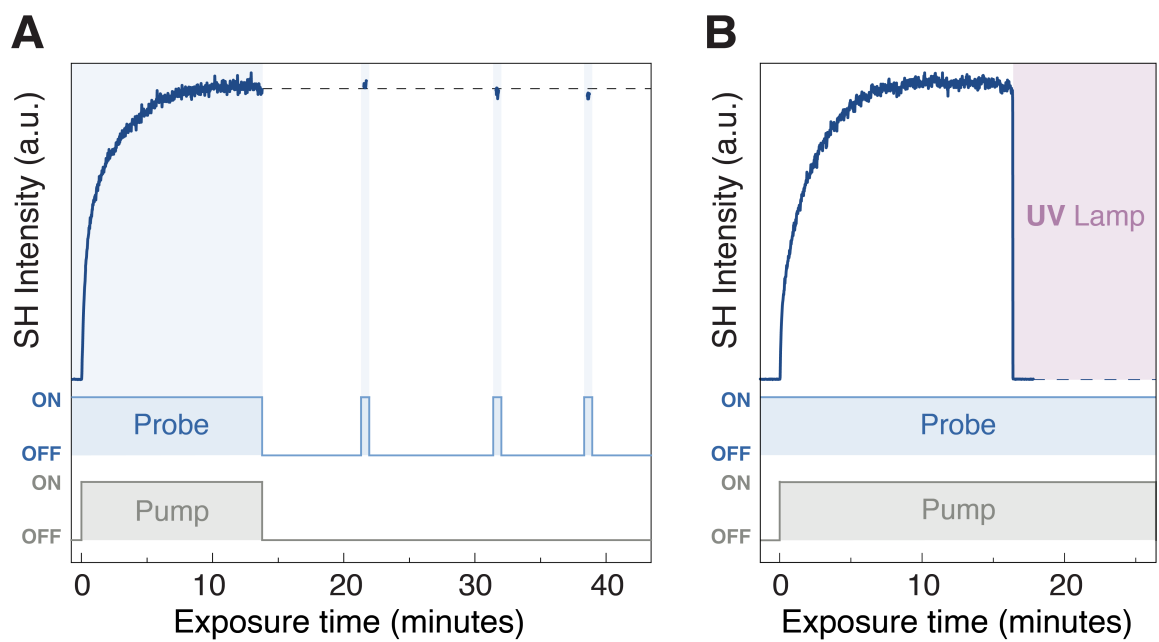

**Figure 3 | Stability of the polar phase. A**, Unperturbed second harmonic intensity as a function of time after exposure to mid-infrared light. After reaching saturation, the pump and probe lights are simultaneously switched off. The second harmonic intensity is briefly and rarely (to avoid any disturbance to the state) sampled at later times by the probe (see also Supplementary S4). After 30 minutes, the second harmonic intensity reduces by only 2%, demonstrating the metastable nature of the induced phase. **B**, Erasure of photo-induced state by above-bandgap illumination. The polar order can be instantaneously erased by exposing the sample to UV light, even with the mid-infrared light still impinging on the sample.

Figure 4 displays time-delay dependent measurements of the same second harmonic signal discussed above. In addition to the time-delay-independent background shown in Fig. 2 – visible as an offset in the traces of Fig. 4A – the second harmonic signal exhibited ultrafast oscillations that resulted from the impulsive inelastic excitation of low-frequency polar modes (see Fig. 4A for three representative measurements taken after 10, 25 and 105 minutes of illumination). The frequency of these oscillations increased visibly with exposure time, as shown more comprehensively in Figure 4B. The detected mode-hardening is reminiscent of the behavior of a ferroelectric soft-mode across a paraelectric to ferroelectric phase transition[19,20] (see inset of Fig. 4B for the example of pressure-induced ferroelectricity in SrTiO$_3$ reproduced from Ref. [[21]]).

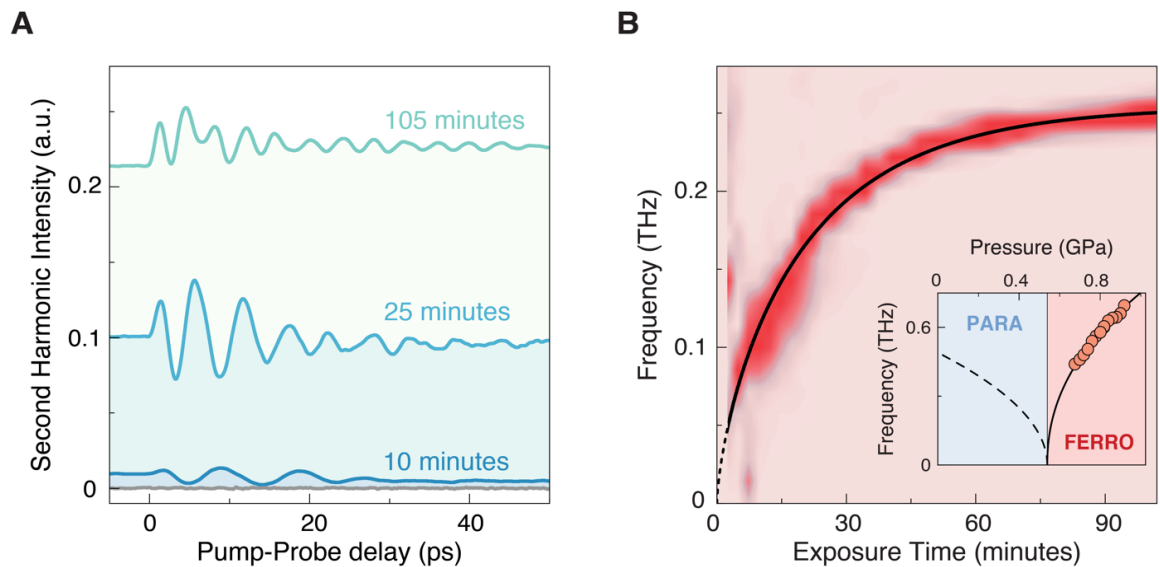

**Figure 4 | Ferroelectric-like mode hardening. A**, Pump-induced ultrafast modulation of the second harmonic intensity as a function of the pump-probe delay for different exposure times. The increasing offset corresponds to the delay-independent second harmonic of Fig. 2B. The grey line represents a measurement taken without mid-infrared pumping. No static second harmonic can be detected. **B**, Fourier transform of the oscillatory traces for different exposure times. The spectra have been normalized for each measurement to better visualize the frequency change and not be affected by the different oscillation amplitudes. The data before minute 5 only contain noise (no second harmonic has grown yet, see Fig. 2A) and, thus, they are not shown. **Inset**, Raman spectrum of strained SrTiO$_3$ as a function of applied pressure (adapted from ref. 21). Experimental data (circles), fit to the data (solid line) and extrapolation to unstrained SrTiO$_3$ (dotted line).

To determine the spatial distribution of the photo-induced polar phase, the experiment was repeated with a collimated 2.2-µm probe beam that was made larger than the photo-excited spot. The second harmonic shadow of the transformed volume was projected onto a CCD camera (Fig. 5A). Two oval bright regions with a dark area in the center were observed (Fig.5B), indicative of an inhomogeneous polar state.

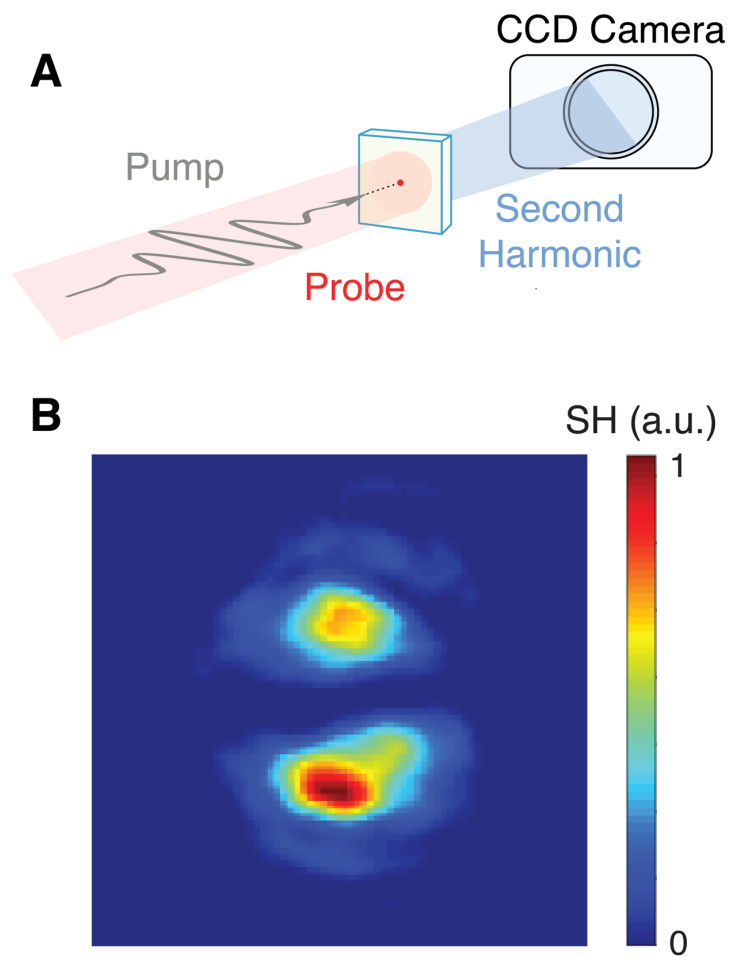

**Figure 5 | Second harmonic spatial dependence. A**, SH shadow imaging setup. A collimated probe beam (much bigger than the pump spot) uniformly illuminates the sample. The emitted second harmonic spatial profile is recorded with a CCD camera (details in Supplementary S7). **B**, Measured spatial dependence of the second harmonic. The color scale is linear.

The inhomogeneity was further evidenced by electrical measurements (Supplementary S9). Gold contacts were deposited onto the sample surface and the optical pump spot was kept much smaller (70 μm diameter) than the distance between the electrodes (1.5 mm).

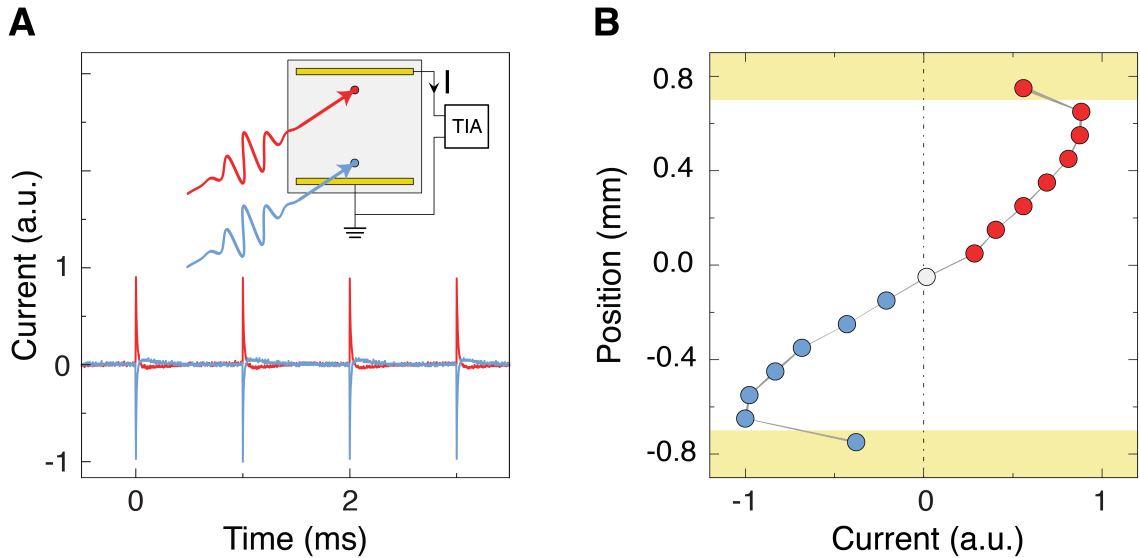

**Figure 6 | Bipolar current pulses in an unbiased sample. A**, Electrical currents induced in SrTiO$_3$ by successive pump pulses (1 KHz repetition rate). The red and blue traces represent measurements taken with the pump impinging closer to the upper and lower contact, respectively. **Inset**, Sketch of the transport measurements setup. Gold contacts were deposited on the sample surface and the electrical response was recorded in short-circuit conditions without any applied bias (TIA = Transimpedence Amplifier; Supplementary S9). **B**, Amplitude of the current pulses as a function of the pump position with respect to the contacts (yellow).

Measurements of the pump-induced electrical response were conducted in under short-circuit conditions, without any applied bias (inset, Fig.6A). Because the optically transformed region was far smaller than the gap between the electrodes, and because the electrodes themselves were not irradiated, the only coupling between the light-induced polar domains and the external circuit was capacitive. The ultrafast response was transmitted to the external circuit, in the form of short bursts of electrical currents at the 1 KHz repetition rate of the pump laser (Fig. 6A, red line).

The sign and amplitude of these current pulses were observed to be dependent on the position of the pump beam. No current pulses were observed when the pump pulse spot was in the middle of the gap (Fig. 6B), and currents of opposite sign (red and blue, Fig.6A) were generated as the spot was moved towards either of the contacts (Fig. 6B).

The observations reported in Figures 5 and 6 are indicative of the creation of *two* oppositely oriented polar domains within the pumped spot, which can be understood as follows.

First, oppositely oriented domains are expected to generate second harmonic light with equal intensities but with a relative $\pi$ shift in their respective optical phases. Hence, as the two second harmonic beams propagate towards the detector, they destructively interfere in the center (Fig. 7C, Supplementary S7), resulting in the observed shadow pattern (Fig. 5B). Second, upon excitation, each of the two domains is expected to draw currents in opposite directions from the electrodes due to the induced polarizations. For a pair of domains aligned at equal distances from the electrodes, no net current is anticipated. However, when these are moved in either direction, the two capacitances would become unequal, drawing a net current in either direction, as observed (Fig. 6B and Fig. 7D).

We next turn to a possible explanation for the formation of this pair of domains. Consider the ionic contribution to electrostriction that nonlinearly couples the optically-pumped phonon mode coordinate $Q_{IR}$ with the strain $\varepsilon$, as detailed in the supplementary information S11. Resonant driving of the $A_{2u}$ optical phonon leads to a transient acoustic deformation of the lattice. According to our estimates, pump peak field strengths of 18 MV/cm, expected to induce ionic oscillations of 5 pm amplitude, result in a dynamical strain of 0.2%. However, if we assume a strain profile with the same gaussian shape of the pump beam, one can readily exclude a direct, strain-induced ferroelectric order, which would result in a single domain with polarization either up or down (Fig. 7A), a homogeneous second harmonic shadow and monopolar current pulses (Supplementary S10). Rather, the coupling between the spatial *gradient* of the optical strain and polarization (Fig.7B), arising from the *flexoelectric* effect[22,23], would result in the creation of two distinct opposing polarizations and would readily explain the data of Figures 5 and 6 (see Fig. 7B-D; Supplementary S8).

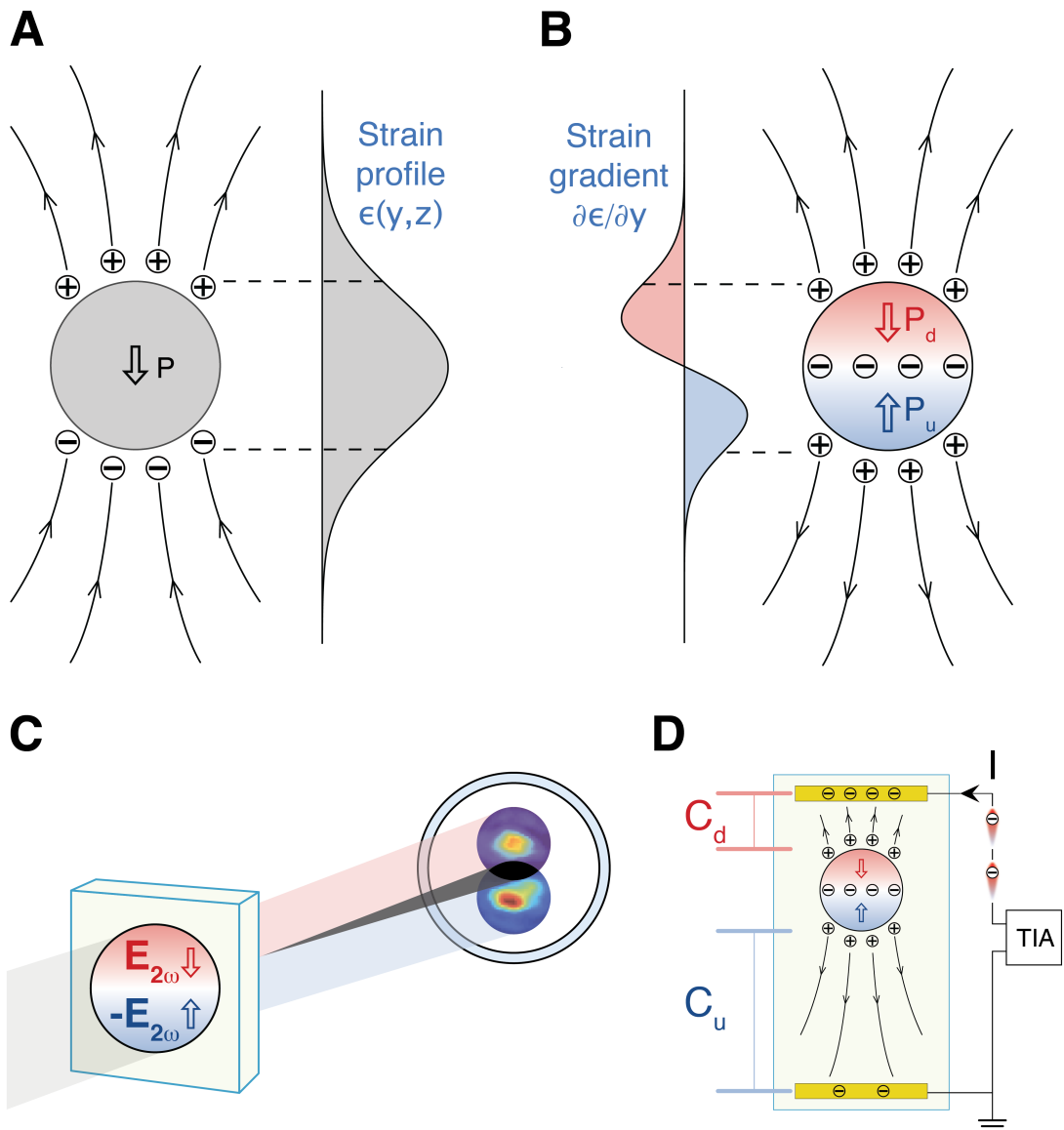

**Figure 7 | Flexo-electric polarization. A**, Phonon pumping results in a strain profile that follows the gaussian profile of the pump beam. The figure depicts how a strain-induced transition to a ferroelectric monodomain state would look like. **B**, Inhomogeneous strain results in two distinct flexo-electric polarizations of opposite polarity that follow the strain *gradient*. **C**, Two opposing polarizations generate second harmonic optical fields in antiphase that interfere destructively on the CCD camera (Supplementary S7). **D**, The position of the flexo-electric domains with respect to the contacts determines the current amplitude and sign (Supplementary S9).

While such a photo-flexo-electric mechanism appears to be a plausible explanation of the source of transiently generated polar domains, this hypothesis does not yet clarify the mechanism that leads to the stabilization of the polarization after the pump is removed. It is quite possible that the ultrafast photo-induced polarization leads to the progressive stabilization of long-range correlations between pre-existing polar nano-regions, a mechanism known for relaxor ferroelectrics and in agreement with existing models for strain-induced ferroelectricity in $SrTiO_3$[24]. Photorefractive effects[25,26], not found in paraelectrics like $SrTiO_3$ but typical of ferroelectric materials, could also contribute to the stabilization of this phase once the local polarizations are created.

These experimental results should motivate optical control experiments that rely on the perturbation of $SrTiO_3$ in functional materials. Because $SrTiO_3$ is used commonly as a substrate for the growth of oxide heterostructures, one could conceive of new ways to drive functional properties at interfaces – including magnetic, electronic and even superconducting states – by photo-induced symmetry breaking. Furthermore, the applicability of the optical control of flexoelectric polarizations extends far beyond the specific case of $SrTiO_3$, since flexoelectricity is allowed in materials of any symmetry[22]. Our findings are also a rare example of real-space symmetry breaking by light, and can thus be considered in the context of other types of light-induced ordering in the presence of fluctuations, such as photo-induced superconductivity well above $T_c$[27,28].

# Supplementary information for "Metastable ferroelectricity in optically strained SrTiO$_3$"

T. F. Nova[1,2], A. S. Disa[1], M. Fechner[1], A. Cavalleri[1,2,3]

[1]Max Planck Institute for the Structure and Dynamics of Matter, 22761 Hamburg, Germany
[2]The Hamburg Centre for Ultrafast Imaging, 22761 Hamburg
[3]University of Oxford, Clarendon Laboratory, Oxford OX1 3PU, UK

## SUPPLEMENTARY SECTIONS:

**S1.** EXPERIMENTAL SETUP
**S2.** SECOND HARMONIC CHARACTERIZATION
　A. SECOND HARMONIC DETECTION
　B. DESCRIPTION OF POLARIMETRY MEASUREMENT AND ANALYSIS
　C. DISCUSSION OF QUADRUPOLE AND SURFACE SHG CONTRIBUTIONS
**S3.** SECOND HARMONIC TEMPERATURE DEPENDENCE
**S4.** POLAR STATE STABILITY UNDER CONTINUOUS PROBING
**S5.** PUMP FLUENCE DEPENDENCE
**S6.** PUMP WAVELENGTH DEPENDENCE
**S7.** SECOND HARMONIC SHADOW IMAGING SIMULATION
**S8.** POLARIZATION DISTRIBUTION FROM PHONON-FLEXOELECTRIC EFFECT
**S9.** DESCRIPTION OF ELECTRICAL MEASUREMENTS
**S10.** COMPARISON OF SINGLE AND DOUBLE DOMAIN MODELS
**S11.** PHONON-INDUCED STRAIN THEORY: MODEL AND FIRST PRINCIPLE CALCULATIONS

## S1. EXPERIMENTAL SETUP

The mid-infrared (mid-IR) pulses used in the experiment were generated by difference frequency generation (DFG) in a GaSe crystal between the signal outputs of two optical parametric amplifiers (OPAs). The OPAs were pumped with 4 mJ pulses, 100 fs long, at 1 KHz repetition rate and 800 nm wavelength delivered from a Coherent Legend regenerative amplifier. The two OPAs were seeded by the same white light continuum (WLC), producing phase locked signal pulses (300 µJ, ~ 80 fs long, independently tunable from 1.2 µm to 1.5 µm). As a consequence, the generated mid-IR transients were carrier-envelope-phase (CEP) stable. The pump pulse frequency was determined by a Michelson Fourier Transform Interferometer. The mid-infrared beam was focused onto the sample using an off-axis parabolic mirror down to 72 µm FWHM. A small fraction of the idler output (2.2 µm, ~ 80 fs long, also CEP stable) of one of the OPAs was used as the probe fundamental for second harmonic generation (SHG). Polarization control of the idler was achieved with a half-waveplate. Pump and probe were kept collinear and at normal incidence. After the sample, a short-pass filter (1500 nm cut-off wavelength) was used to isolate the second harmonic signal (1.1 µm) from the fundamental (2.2 µm); the pump was entirely blocked by the material. At need, a polarizer was employed to select specific components of the SH electric field. The SH photons were detected using an InGaAs Femtowatt Photoreceiver from Newport (800-1700 nm sensitivity). The output of the diode was sent to a lock-in amplifier. In the shadow imaging measurements, the diode was replaced by a Thorlabs DCC1545M CMOS camera (see Supplementary S7).

The schematic representation of the setup is shown in Fig. (S1.1).

The sample used in the experiment is a bulk single crystal of (110)-cut $SrTiO_3$, 50 micrometers thick. The temperature was controlled with a cold finger cryostat. Two feed-throughs connected low-noise cables to external circuitry for electrical measurements at low temperatures (see Supplementary S9).

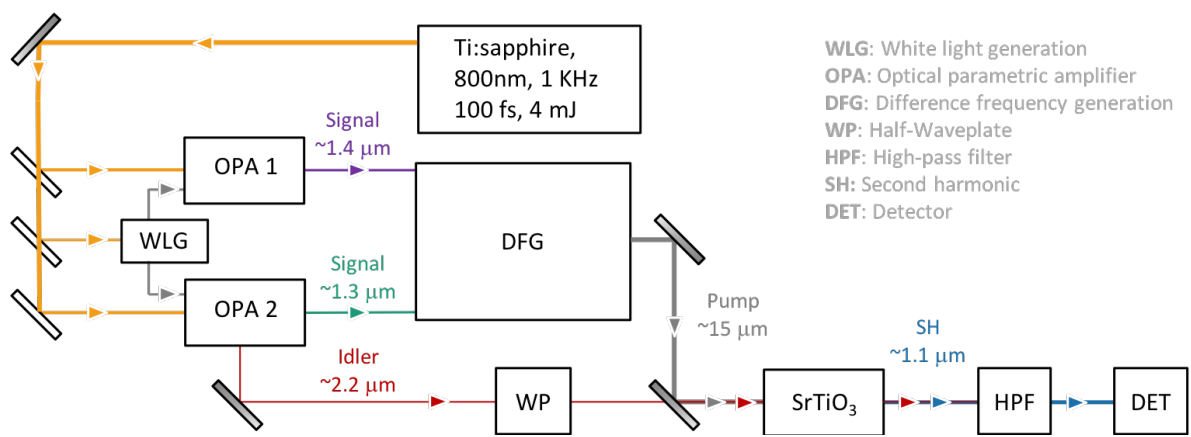

**Figure S1.1 | Experimental setup.**

## S2. SECOND HARMONIC CHARACTERIZATION

### A. SECOND HARMONIC DETECTION

To ensure that only second harmonic photons reached the detector, we employed a short-pass filter, which blocks the fundamental beam (2.2 µm). A spectrometer was used to confirm the wavelength of the detected photons (1.1 µm). The quadratic power dependence of the second harmonic intensity (Fig. S2.1) revealed that the signal on our detector was only due to two-photon processes, as expected (and not to higher order processes or fundamental leakage) [13].

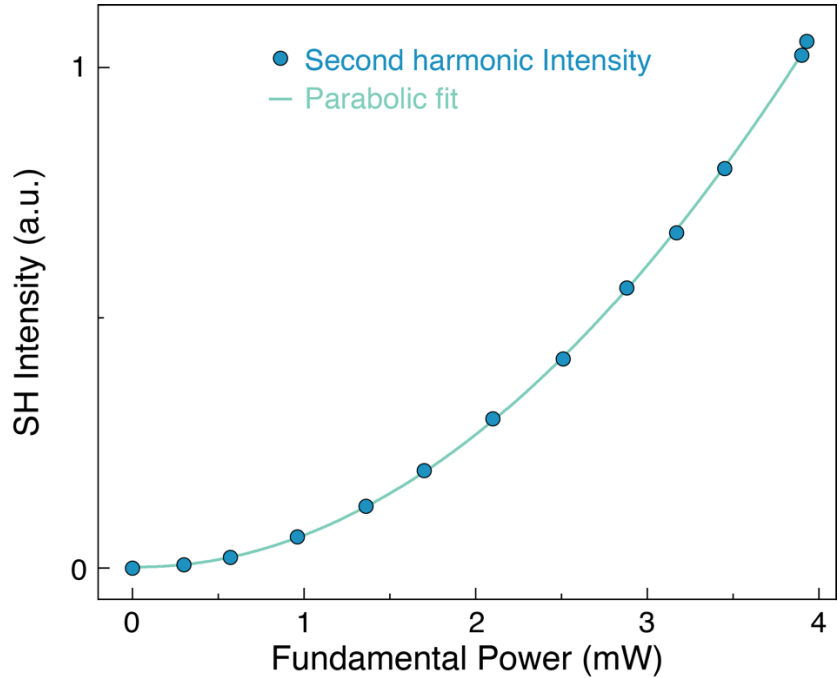

**Figure S2.1 | Second harmonic fluence dependence.** The second harmonic intensity depends quadratically on the input power of the fundamental, thus confirming the nonlinear nature of the generation mechanism.

### B. DESCRIPTION OF POLARIMETRY MEASUREMENT AND ANALYSIS

Within the dipole approximation, second harmonic generation in crystals is described phenomenologically by the relationship [13],

$$P_i^{(2\omega)} = \sum_{jk} \chi_{ijk}^{(2)} E_j^{(\omega)} E_k^{(\omega)}, \tag{S2.1}$$

where $E^{(\omega)}$ is the incident electric field at frequency $\omega$, $P^{(2\omega)}$ is the induced polarization in the crystal at frequency $2\omega$, and $\chi^{(2)}$ is the third rank nonlinear susceptibility tensor. The indices *i,j,k* designate the components along (*x,y,z*). By symmetry, $\chi^{(2)}$ and hence, SHG may only be non-zero in crystals with no center of inversion. There are 21 non-centrosymmetric crystal point groups; however, only 10 are polar – that is, they carry a spontaneous polarization along a unique axis. The form of the $\chi^{(2)}$ tensor (i.e. which elements are non-zero) is dictated by the point group symmetry of the crystal.

Bulk SrTiO$_3$ is centrosymmetric in equilibrium at all temperatures, and below 105 K has a $D_{4h}$ (4/*mmm*) point group symmetry. The detection of finite SHG after pumping SrTiO$_3$ with mid-infrared pulses (Fig. 2, main text) demonstrates that SrTiO$_3$ is driven into a new phase lacking inversion symmetry. In order to determine if the transformed SrTiO$_3$ is polar, we examine the form of $\chi^{(2)}$, whose different elements can be accessed by rotating the polarization of the incident electric field and analyzing the polarization of the SHG, as shown in Fig. 2A of the main text.

In our experimental geometry, the incident probe light propagates normal to the surface of a SrTiO$_3$ (110) crystal. We label the propagation direction *z* = [110] and the electric field is polarized in the plane consisting of the *x* = [001] and *y* = [1-10] axes (in pseudocubic notation). It is convenient to rewrite Eq. (S2.1) in reduced form,

$$P_i^{(2\omega)} = \varepsilon_0 d_{ij} \widetilde{E^2}_j^{(\omega)}, \tag{S2.2}$$

where *j* indexes the components of the vector $\widetilde{E^2}^{(\omega)} = (E_x^2, E_y^2, E_z^2, 2E_yE_z, 2E_xE_z, 2E_xE_y)$, $d_{ij}$ is the reduced nonlinear susceptibility tensor, and $\varepsilon_0$ is the vacuum permittivity. In our polarimetry measurement, we rotate the linear polarization of the incident light in the *xy* plane with a half-waveplate and detect the transmitted SHG polarized either along *x* or *y* with an analyzer. As a function of the probe polarization angle $\theta$ relative to the *x* axis, Eq. (S2.2) becomes,

$$P_x(2\omega) = \varepsilon_0 E_0^2(\omega)[d_{11}\cos^2\theta + d_{12}\sin^2\theta + 2d_{16}\cos\theta\sin\theta] \tag{S2.3}$$
$$P_y(2\omega) = \varepsilon_0 E_0^2(\omega)[d_{21}\cos^2\theta + d_{22}\sin^2\theta + 2d_{26}\cos\theta\sin\theta],$$

for *x* and *y* analyzer polarizations, respectively. Ultimately, we detect the SHG intensity from the pumped SrTiO$_3$ state, which leads to the "flower plots" shown in Fig. 2C in the main text. The intensities $I_x(2\omega) \propto |P_x(2\omega)|^2$ and $I_y(2\omega) \propto |P_y(2\omega)|^2$ are simultaneously fit using Eqs. (S2.3) with free parameters $E_0$, $d_{11}$, $d_{12}$, $d_{16}$, $d_{21}$, $d_{22}$, and $d_{26}$. Note that the elements above are given in the laboratory frame, and knowledge of the symmetry axes in the pumped SrTiO$_3$ phase is necessary to write $d_{ij}$ in the crystal frame.

For the fits shown in the main text, the values of $d_{ij}$ are given in Table (S2.1). One can discern two important aspects from the table: first, the largest element by almost a factor of two is $d_{22}$; and second, the magnitude of $d_{11}$ is less than 10% of $d_{22}$ and those of $d_{12}$ and $d_{26}$ are less than ~1% of $d_{22}$. We thus presume that the primary symmetry axis in the pumped STO state is along *y* = [1-10] and that the $d_{11}$, $d_{12}$ and $d_{26}$ elements may be zero by symmetry. In the new crystal frame, these constraints correspond to requiring the nonlinear susceptibility tensor to have non-zero elements $d_{15}$, $d_{31}$, $d_{33}$. Only nine point groups (besides the triclinic $C_1$) fulfill this requirement: $C_s$, $C_2$, $C_{2v}$, $C_4$, $C_{4v}$, $C_3$, $C_{3v}$, $C_6$, and $C_{6v}$. All of the possible point groups are polar, and of these, five ($C_s$, $C_2$, $C_{2v}$, $C_4$, $C_{4v}$) are subgroups of the equilibrium point group of SrTiO$_3$ ($D_{4h}$). A group-subgroup relation between the initial and transformed SrTiO$_3$ would necessarily hold if the pump-induced phase transition is considered second order and/or displacive in nature. The four groups $C_2$, $C_{2v}$, $C_4$, or $C_{4v}$ cannot be differentiated by polarimetry in our experimental geometry. The fitted values for $d_{ij}$ assuming one of these groups are given in Table (S2.1). The fit quality is similar to that of the $C_s$ group, for which all accessible elements are non-zero (see Fig. (S2.2)).

The symmetry analysis demonstrates that the pump-induced state of SrTiO$_3$ is polar with a spontaneous polarization oriented along the [1-10] direction. Its point group symmetry is consistent with that determined for $^{18}$O-isotope substituted and Ca-substituted ferroelectric SrTiO$_3$ ($C_{2v}$) [14,15]. We note that the polarimetry results are also consistent with other polar point group symmetries; however, the important conclusion that the SrTiO$_3$ develops a spontaneous polar order following mid-IR exposure remains valid.

**Table S2.1 | Fit values of the reduced nonlinear susceptibility tensor (defined in Eq. (S2.2)) corresponding to the data shown in Fig. 2C in the main text.** The components of $d_{ij}$ are given in both the lab and crystal frame (see text). The fit values are shown for the case in which the point group symmetry is assumed to belong to $C_1$ or $C_s$, as well as the case in which the symmetry is assumed to be $C_2$, $C_{2v}$ $C_4$, or $C_{4v}$. In the latter case, three elements are zero by symmetry.

| $d_{ij}$ (lab frame) | $d_{ij}$ (crystal frame) | Fit value ($C_1$,$C_s$) | Fit value ($C_2$,$C_{2v}$ $C_4$,$C_{4v}$) |
|---|---|---|---|
| $d_{11}$ | $d_{11}$ | 0.462 | 0 |
| $d_{12}$ | $d_{12}$ | 0.047 | 0 |
| $d_{16}$ | $d_{15}$ | 3.648 | 3.417 |
| $d_{21}$ | $d_{31}$ | 1.284 | 1.195 |
| $d_{22}$ | $d_{33}$ | 5.351 | 5.804 |
| $d_{26}$ | $d_{35}$ | -0.056 | 0 |

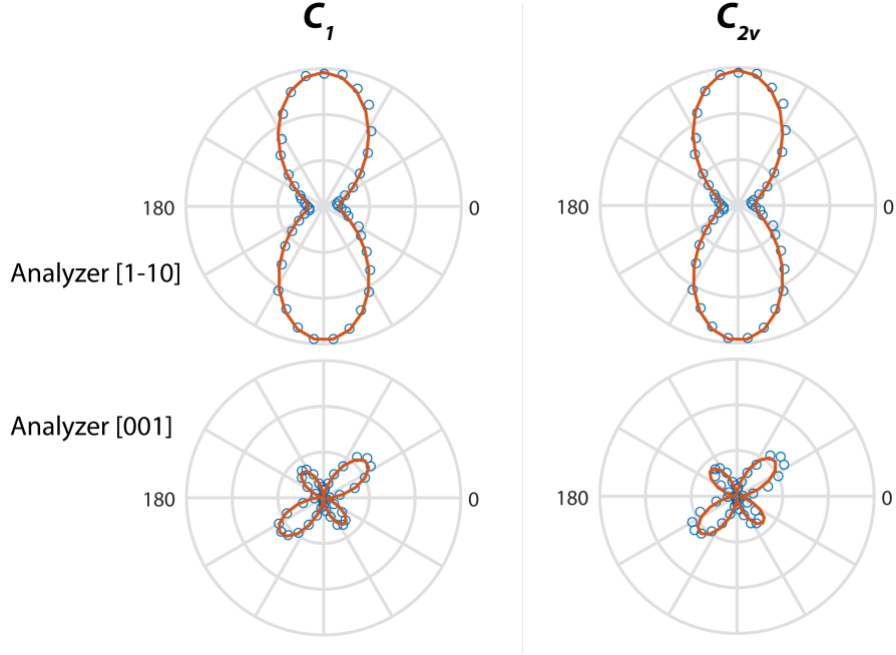

**Figure S2.1 |** Comparison between fits of SHG polarimetry using Eqs. (S2.3) and the values from Table (S2.1) assuming $C_1$ (or $C_s$) point group symmetry (left) and $C_{2v}$ (or $C_2$, $C_{4v}$, or $C_4$) point group symmetry (right).

## C. Discussion of quadrupole and surface SHG contributions

In general, centrosymmetric media may also exhibit SHG arising from bulk electric quadrupole and surface electric dipole contributions. The effective second harmonic polarization due to the electric quadrupole term can be written as [16],

$$P_i(2\omega) = \sum_{jkl} \Gamma_{ijkl} E_j(\omega) \nabla_k E_l(\omega), \quad (S2.4)$$

where $\Gamma_{ijkl}$ is the fourth rank nonlinear quadrupole susceptibility tensor, and the indices refer to the laboratory frame. In our experimental geometry, we assume a transverse plane wave propagating normal to the $z$ = [110] direction in the crystal, so the electric field is given by,

$$\vec{E}(\vec{r}, t) = E_0 e^{i(\kappa z - \omega t)}[\cos\theta\,\hat{x} + \sin\theta\,\hat{y}], \quad (S2.5)$$

with the field amplitude $E_0$, the wavenumber $\kappa$ = $n\omega/c$, and the polarization angle $\theta$ with respect to the $x$ axis. It can be seen from inserting the expression for $E$ from Eq. (S2.5) into Eq. (S2.4) that the only accessible terms in this geometry are those with $k = z$ and $j,l = x$ or $y$. Furthermore, we analyze only the emitted SHG light polarized in the $xy$ plane, so that the index $i = x$ or $y$. Thus, the experimentally accessible terms of Eq. (S2.4) involve components of the quadrupole susceptibility of the form $\Gamma_{ij3l}$ with $i,j,k = x$ or $y$.

For the equilibrium tetragonal $D_{4h}$ symmetry of SrTiO$_3$, there are only 21 non-zero elements of $\Gamma_{ijkl}$, nine of which are independent. The tensor elements are given by [16],

$$\Gamma_{ijkl} = a_1\delta_{ijkl} + a_2(\delta_{ij}\delta_{kl} + \delta_{il}\delta_{jk}) + a_3\delta_{ik}\delta_{jl} + a_4\delta_{ij}\delta_{zz} + a_5\delta_{il}\delta_{jk(z)} + a_6\delta_{ik}\delta_{jl(z)} + a_7\delta_{jl}\delta_{ik(z)} + a_8(\delta_{ij(z)}\delta_{kl} + \delta_{il(z)}\delta_{jk}) + a_9\delta_{ijkl(z)},\qquad(S2.6)$$

where $\delta$ is the Kronecker delta, and $a_i$ are constants. Terms of the form $\delta_{ij}\delta_{kl}$ imply $i,j \neq k,l$ and the label $(z)$ signifies $\delta_{ij(z)} = 1$ when $i=j=z$. From Eq. (S2.6), one can see that $\Gamma_{ijkl}$ is only non-zero in cases when each index appears an even number of times. This requirement is incompatible with the accessible terms discussed above. Hence, the electric quadrupole contribution to our measured SHG signal is zero in our experimental geometry.

The crystal surface provides an additional source of SHG, even in centrosymmetric materials, due to the inherent inversion asymmetry at the vacuum-bulk interface. In SrTiO$_3$ (110), the point group symmetry reduces from $D_{4h}$ to $C_s$ at the surface, with the polar axis ∥ $z$. For this symmetry, the dominant accessible terms in our experimental geometry ($d_{21}$, $d_{22}$, $d_{16}$) are not included in the non-zero tensor elements for the surface contribution to $\chi^{(2)}$, though other terms ($d_{11}$, $d_{12}$, $d_{26}$) are. In addition, we note that before pumping the sample with mid-IR light, no SHG was detectable. Thus, we conclude that the surface contribution to the SHG signal is negligible in our experimental geometry.

The above analysis indicates that both the bulk electric quadrupole and surface sources of SHG are small or not observable in our experimental geometry. Therefore, the observed SHG signal in the pumped SrTiO$_3$ state arises from the electric dipole term and is a reliable indication of bulk symmetry breaking.

## S3. SECOND HARMONIC TEMPERATURE DEPENDENCE

In Fig. (S3.1) we show the temperature dependence of the saturation value of the second harmonic signal of Fig. 2B of the main text. At each temperature, we induced the polar state and waited until saturation was reached. Afterwards, we erased the state by briefly exposing the sample to UV light and changed the temperature. We repeated the procedure starting from 4 K to room temperature. As can be seen from the inset of Fig. (S3.1), even at room temperature we could observe a mid-IR-induced second harmonic signal. The signal did not seem to be affected by the cubic-to-tetragonal structural transition naturally occurring in $SrTiO_3$ at 105 K [24]. On the other hand, the size of the photo-induced effect seemed to closely follow the behavior of the static dielectric function [23].

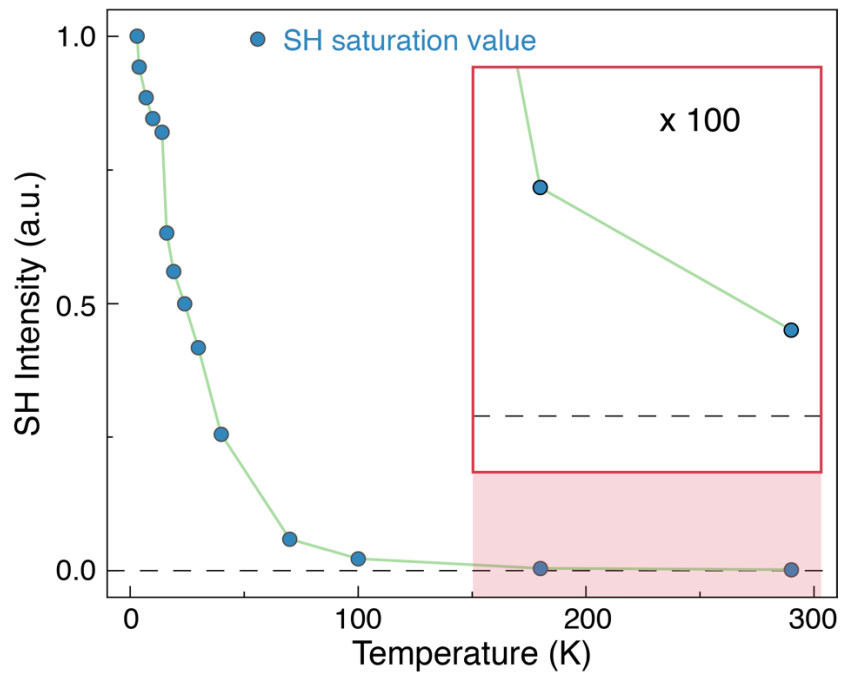

**Figure S3.1 | Second harmonic temperature dependence.** The blue dots represent the saturation value of the second harmonic at each temperature. The inset shows a 100 times magnification of the high temperature results.

## S4. POLAR STATE STABILITY UNDER CONTINUOUS PROBING

In the main text (Fig. 3A) we have shown how the photo-induced state has a lifetime of several hours even after the pump has been switched off. In those measurements, the second harmonic value was sampled only briefly and rarely to perturb the polar order as little as possible. In Fig. (S4.1), we show the behavior of the polar state after removing the pump while being continuously sampled. In this case, the lifetime strongly reduces to only few tens of minutes, thus implying that the probe photons destructively interact with the phonon-induced polar state. This observation can be understood as follows. In Fig. 3B of the main text we show how exposure to above-band-gap photons (that is the creation of photo-injected free carriers) efficiently destroys the polar order [8]. For 1.1 μm second harmonic photons (far smaller than the gap) the absorption probability is heavily reduced, yet finite. Thus, over a scale of minutes (and hundreds of thousands of pulses), the probe will also be detrimental for the state. We repeated the same experiments for 1.6 μm and 0.8 μm probe photons (0.8 μm and 0.4 μm second harmonic photons, respectively). In those cases, it became increasingly harder to observe second harmonic growth (still sizable for 1.6 μm light, while undetectable for 0.8 μm photons). These results highlight the importance of the choice of the probe photon energy to sample the polar state without perturbing it.

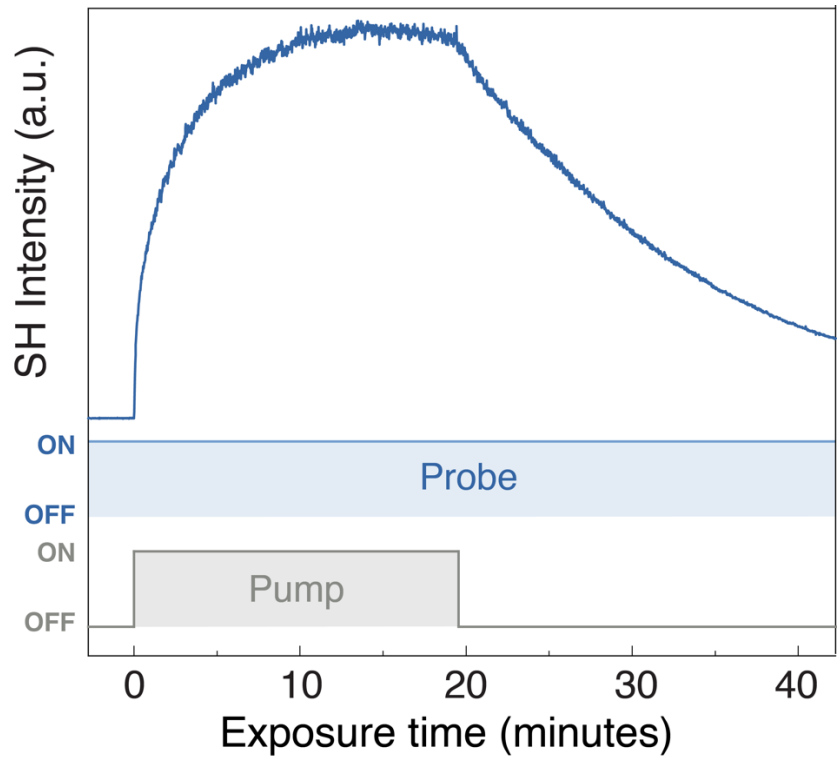

**Figure S4.1 | Second harmonic intensity decay under continuous sampling.** The probe beam samples the second harmonic signal for the entire time of the experiment, while the pump is on only for the first 20 minutes

## S5. PUMP FLUENCE DEPENDENCE

In Fig. (S5.1) we show the pump fluence dependence of the second harmonic saturation value. For each fluence, we measured the second harmonic growth and waited until saturation was reached. Subsequently, we erased the state by briefly exposing the sample to UV light and repeated the experiment with a higher fluence. The fluence dependence reveals two important pieces of information: first, a minimum threshold is needed to initiate the reported second harmonic growth (for a 15 μm pump, the threshold is 2.7 mJ/cm²). Second, a saturation behavior is observed with fluence. The data in Fig. (S5.1) can be fit by the function

$$f(F) = \alpha\left(1 - e^{-\chi(F-\beta)}\right), \tag{S5.1}$$

where $F$ is the fluence, and $\alpha$, $\beta$ and $\chi$ are the saturation value, the threshold fluence and the photo-susceptibility, respectively. The photo-susceptibility represents a measure of how efficiently the pump energy is used to create the polar state.

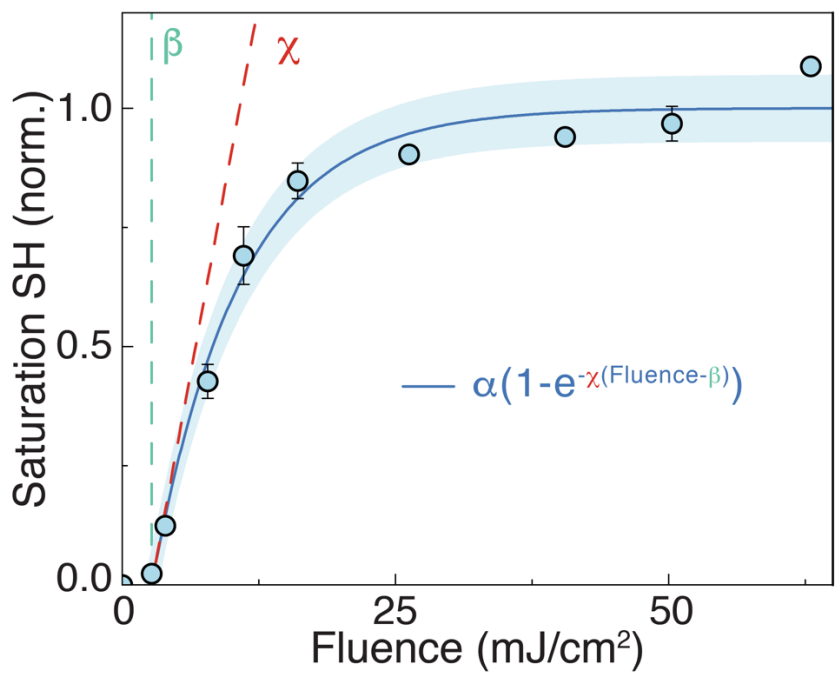

**Figure S5.1 | Pump fluence dependence.** We display the second harmonic saturation value as a function of pump fluence for a 15 μm pump wavelength.

## S6. PUMP WAVELENGTH DEPENDENCE

To confirm the vibrational origin of the reported effect, we systematically varied the wavelength of the pump pulses. For each wavelength (photon energy), we repeated the fluence dependence measurement of Fig. (S5.1). As a measure of how efficiently the growth mechanism can be initiated we extracted the photo-susceptibility $\chi$ (as defined in Eq. (S5.1)) for every pump wavelength. The results of the inset of Fig. 2A (main text) show that the photo-susceptibility maximum is reached for pump photon energies close to the resonance of the $A_{2u}$ phonon. For wavelengths approaching the optical gap (3.2 eV), the effect is severely reduced if not completely absent.

## S7. SECOND HARMONIC SHADOW IMAGING SIMULATION

We performed beam propagation simulations to confirm that the shadow image shown in Fig. 5B in the main text arises from the interference of two anti-phase sources of SHG, as in the case of oppositely poled ferroelectric domains. Simulations were carried out using Zemax OpticStudio, and we follow an approach similar to the split mirror design of Ref. [17]. The input beam is Gaussian with a 1.1 µm wavelength and 5 mm waist. We mimic the effect of two anti-phase domains by using two non-absorbing, linear, dielectric elements, which are stacked bisect the beam in the direction of propagation. One element has thickness $\lambda/n$ and the other $\lambda/2n$. A 70 µm diameter circular aperture was placed in front of the elements to define the pumped region. The input beam is weakly focused by a long focal length lens ($f$ = 1200 mm) and the aforementioned structure is placed close to the focus. The far-field image is sampled 100 mm from the aperture. The result is shown in Fig. (S7.1).

The experimental image was produced in a similar geometry, with the sample replacing the split structure. The probe beam in the experiment had a 2.2 µm wavelength and the 1.1 µm second harmonic was generated in the ~70 µm pumped region. The focal length of the input lens in the experimental set up was f = 1200 mm (used to compensate for a slight divergence of the incoming beam and to avoid beam clipping on the mid-infrared focusing optics). The SHG image was recorded with p polarization of the probe and analyzer on a Thorlabs DCC1545M CMOS camera placed ~100 mm from the sample. We note that the simulation considers only the linear propagation of optical beams and does not take into account any nonlinear effects such as self-focusing, which may play a role in the slight discrepancy between simulation and experimental results. The minor asymmetry in the simulated lobes is due to an offset (~1 micron) between the circular aperture and the boundary between the two optical elements.

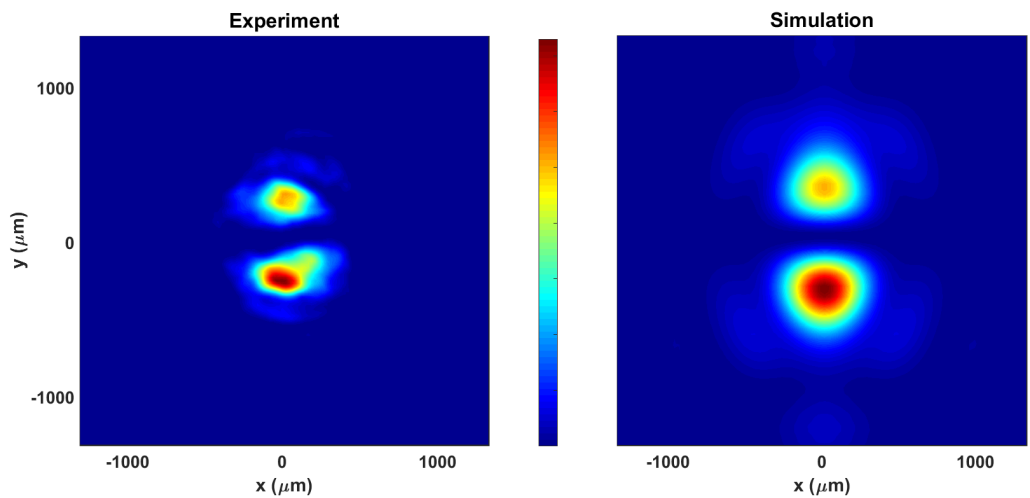

**Figure S7.1 | Shadow imaging, experiment vs simulations.** Comparison between measured (left) and simulated (right) SHG shadow image. The color scale is linear.

## S8: Polarization distribution from phonon-flexoelectric effect

The direct flexoelectric effect describes the polarization induced by a strain gradient, which can be expressed as

$$P_i = \sum_{jkl} f_{ijkl} \nabla_j \epsilon_{kl}, \tag{S8.1}$$

where $P_i$ is the polarization, $\epsilon_{ij}$ is the strain tensor, and $f_{ijkl}$ is a fourth-rank flexoelectric tensor. From the phonon-induced stress given by Eq. (S11.5) which is quadratic in the phonon amplitude $Q$, and the linear relationship between the pump electric field and $Q$, we can determine the spatial profile of the induced strain $\epsilon_{ij}(x,y,z)$ on the sample. We assume a Gaussian pump beam, whose intensity is given by

$$I(x,y,z) = I_0 e^{-\frac{(4\ln 2)x^2}{w^2}} e^{-\frac{(4\ln 2)y^2}{w^2}} e^{-\frac{z}{\delta}}, \tag{S8.2}$$

with $I_0$, a constant, $w$, the full width at half maximum (FWHM) of the pump spot, and $\delta$, the penetration depth of the pump. The coordinate system is the same as that used in Sec. (S2B) (lab frame: $x = [001]$, $y = [1\text{-}10]$, $z = [110]$). Combining Eqs. (S8.2), (S11.10), and (S11.6) to compute (S8.1), gives

$$P_i = \sum_{jkl} I_0 s_{ijkl} q_{ijj} a^2 e^{-\frac{(4\ln 2)(x+y)^2}{w^2}} e^{-\frac{z}{\delta}} \left[ (f_{i1kl} x + f_{i2kl} y)\left(-\frac{8\ln 2}{w^2}\right) + f_{i2kl}\left(-\frac{1}{\delta}\right) \right], \tag{S8.3}$$

in which $s_{ijkl}$ and $q_{ijj}$ are the elastic compliance tensor of SrTiO$_3$ and the coupling constant, respectively, from Eq. (S11.6).

Experimental values for $f_{ijkl}$ in SrTiO$_3$ are available only in the cubic phase, in which there are three independent elements of the form $f_{iiii}$, $f_{iijj}$, and $f_{ijij} = f_{ijji}$. Phenomenologically, it is found that the flexoelectric effect scales with the dielectric constant of the material; so we adapt the cubic values to the low temperature phase by including the strong dielectric anisotropy of SrTiO$_3$ through the relationship [21], $f'_{ijkl} = \sum_m \frac{\chi_{im}}{\chi_c} f_{mjkl}$, where $\chi_{im}$ is the linear electric susceptibility tensor in the low temperature phase and $\chi_c$ is the susceptibility in the cubic phase. The values for $f_{ijkl}$ are taken from Ref. [22] and the values for $\chi_{im}$ from Ref. [23]. The resulting polarization field is shown in Fig. (S8.1), demonstrating the development of two domains with opposite polarization.

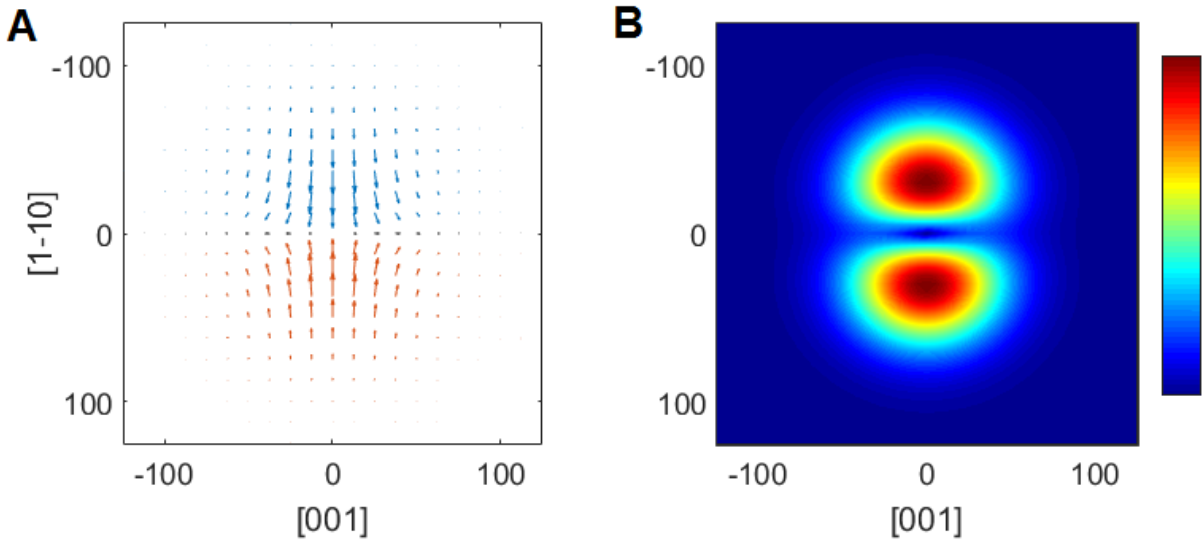

**Figure S8.1 | Phono-flexoelectric polarization. (A)** Calculated vector field and **(B)** magnitude of the polarization from the phonon-induced flexoelectric effect, Eq. (S8.3)**.**

## S9. DESCRIPTION OF ELECTRICAL MEASUREMENTS

The electrical measurements shown in Fig. 6 in the main text were carried out in a $^4$He flow cryostat on a 50μm thick STO (110) crystal. Two 4 mm wide Ti/Au contacts were deposited on the sample using e-beam evaporation defined by an aluminum shadow mask, separated by a distance of 1.5 mm along the [1-10] direction (as depicted in the inset of Fig. 6A in the main text). One contact was directly connected to ground, while the other was connected to the input of a Model DLPCA-200 Femto transimpedance amplifier, which shares the same ground. The transimpedance gain for the measurements shown was V/A = $10^9$, which provides a -3 dB bandwidth of 50 kHz and a rise time of 7μs. The output signal from the amplifier was acquired using a digital oscilloscope with a 2 GS/s sampling rate. In such a configuration, the oscilloscope measures a voltage proportional to the short-circuit current flowing between the two contacts.

Upon the arrival of a mid-infrared pulse, we observe a sharp pulse of current, a fast decay, and subsequent slow oscillation and decay (Fig. 6A in the main text). When the pump light is blocked, no current is detectable, even under the highest gain settings (V/A = $10^{11}$). When the pump light is allowed to impinge upon the sample in the region between the two contacts, the transient current response is observable with the same repetition rate as the pump (1 kHz). The rise time of the current pulse matches the rise time of the transimpedance amplifier, the initial fast decay occurs on a time scale of around 10-20 μs, and the slow decay around 0.2-0.5 ms. We note that this behavior may reflect the response of the amplifier to a sudden impulse faster than its rise time rather than the true time structure of the transient current. Nevertheless, such a signal undoubtedly implies the existence of a short-circuit (zero-bias) photocurrent that appears due to the pump.

To measure the spatial dependence of the current pulses, the output of the transimpedance amplifier was connected to a lock-in amplifier triggered at 1 kHz. The sample was moved vertically in 100 μm steps from one contact to the other, and the output voltage of the lock-in was recorded at each position.

We emphasize here that the pump photon energy in our experiment is far below the band gap of SrTiO$_3$ in equilibrium ($E_{pump}/E_{gap} \approx 0.02$), so the observed current is unlikely to originate from the conventional photovoltaic effect [18,19]. Moreover, since the size of the pump pulse (70μm FWHM) is much smaller than the distance between the contacts, the influence of band bending at the metal/dielectric (contact/sample) interface can be ruled out, as the typical depletion width for an Au/SrTiO$_3$ junction is on the order of 0.1-1 μm [20]. Hence, we attribute the observed current to the creation/enhancement of a pump-induced polarization akin to the switching current of a poled ferroelectric. The mechanism is discussed further in the main text.

## S10: COMPARISON OF SINGLE AND DOUBLE DOMAIN MODELS

Here, we discuss in more detail the experimental signatures of the single vs. double polar domain structure. As mentioned in the main text, the coupling of a pump-driven phonon with strain could result in two different polarization profiles. One would arise from a direct piezoelectric-type coupling, in which $P \propto \epsilon$, and the other would arise from a flexoelectric coupling of $P \propto \nabla\epsilon$. Since the strain is proportional to the Gaussian intensity profile of the pump, the aforementioned couplings imply either a single polar domain or two polar domains with opposite sign, respectively. For each case, we consider two experimental observations: the spatial dependence of pump-induced short-circuit current pulses (Fig. (S10.1A)) and the SHG image produced from a collimated probe (Fig. (S10.1D)).

First, we consider the measurement of current pulses. We assume the pump-induced current results from the capacitive coupling between the domain boundary and the contacts, and the location of the pump spot relative to the contacts changes the relative capacitance to the top ($C_d$) and bottom ($C_u$) contacts (see Sec. (S9) and Fig. 7D in the main text). A single pump-induced domain will have an average boundary charge density $+\sigma_b$ on the upper half of the domain and $-\sigma_b$ on the bottom half. A compensating (screening) charge density will form as a consequence of the domain structure with negative charge on the top contact and positive charge on the bottom contact, driving a current in the process (whose polarity we will call positive). Changing the capacitances $C_d$ and $C_u$ will affect the magnitude of the screening charge on each contact, while the sign of the charges is fixed due to the domain structure. Hence, the polarity is always positive (Fig. (S10.1B)). In the case of two domains, since the domains have opposite sign, the charge density will have of the same sign throughout the entire boundary of the pumped region ($+\sigma_b$). The screening charges that arise will, therefore, be negative on both contacts, but their relative magnitude will be dictated by the ratio of $C_d$ to $C_u$, which will also determine the direction of current flow. Since $C \propto \frac{1}{d}$, where $d$ is the distance between the domain edge and the contact, $C_d$ decreases and $C_u$ increases when the pump spot is closer to the top contact and vice versa for the bottom contact. Hence, the relative charge density and polarity of current will be negative or positive when the pump spot is closer to the top or bottom contact, respectively, as observed in the experiment (Fig. (S10.1C) and Fig. 6A in the main text).

Now, we consider the SHG shadow imaging. In the experiment, a collimated probe with wavelength $\lambda$ = 2.2 µm impinges on the pumped region whose size is much smaller than the probe size. The pumped region acts as a source of SHG with $\lambda^{(2)}$ = 1.1 µm and will emit light with divergence angle $\theta \sim \frac{\lambda^{(2)}}{D}$, where $D$ is the size of the source domain. As shown in Fig. (S10.1E) a single domain acts as a source of SHG, emitting light with uniform phase and creating a circular far-field intensity profile, in analogy with the single-slit diffraction pattern. In the double domain case, the pump creates two sources of SHG, each roughly half the size of the pumped region. The sign of the polarization determines the phase of the emitted radiation, so that the SHG from the upper and lower domains are π phase-shifted relative to each other. Due to the divergence of the emitted SHG, the light from the two domains overlap and interferes destructively on the screen, creating a far-field image with an empty central region and intense side lobes, as shown in Fig. (S10.1F) and simulated in Sec. (S7).

Thus, from general arguments, the model of two oppositely poled domains caused by a phonon-flexoelectric coupling is able to reproduce our experimental results. The model of a single domain from direct phonon-strain coupling would produce contradictory observations, ruling it out as a realistic model.

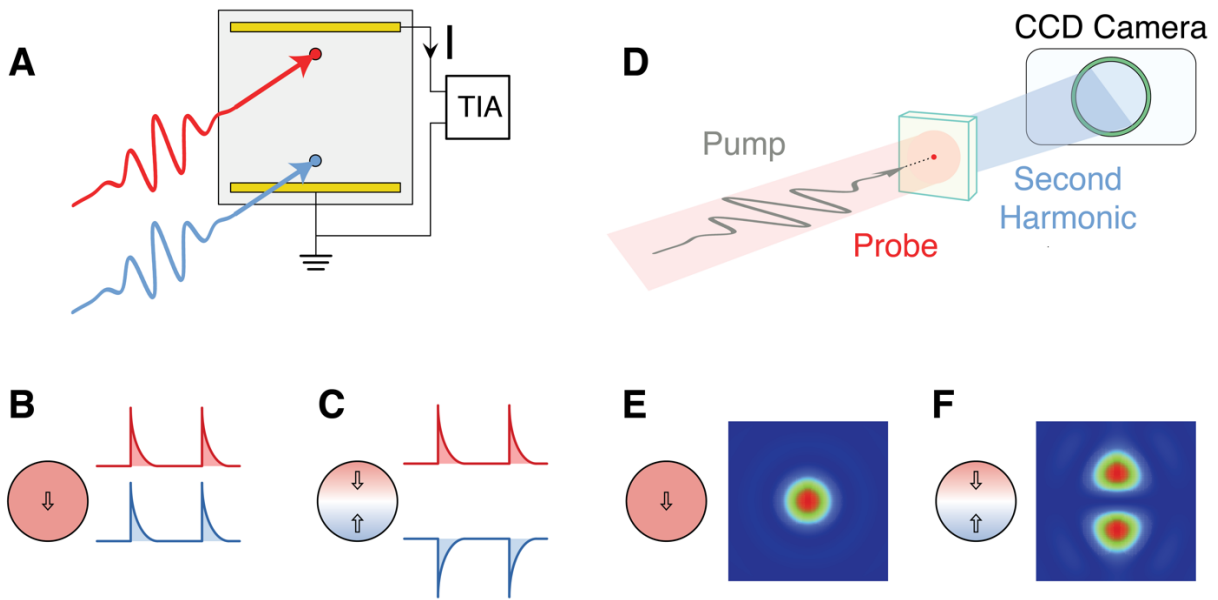

**Figure S10.1 | Comparison of predicted observations for single and double domain models. (A)** Schematic of the measurement of the spatial dependence of current pulses (TIA = transimpedance amplifier). **(B)** A single domain is expected to produce currents with the same polarity, while **(C)** two oppositely poled domains would produce currents with opposite polarity. **(D)** Schematic of the SHG shadow imaging measurement. **(E)** A single domain is expected to produce a single uniform spot, while **(F)** two oppositely poled domains would produce an interference pattern with two lobes and an empty central region. The experimental observations shown in the main text match the double domain model predictions (C) and (F).

## S11. PHONON-INDUCED STRAIN THEORY: MODEL AND FIRST-PRINCIPLES CALCULATIONS

In this section, we discuss the theory of phonon-induced strain based on the expansion of the free energy. Furthermore, we compute a quantitative estimate of its size for SrTiO$_3$ based on first principle calculations.

A full description of the elastic and structural properties of materials requires the expansion of the free energy in both strain and phonon variables [1]. We write this expansion in a compact form as

$$E = \frac{1}{2}\sum_i c_{ii} \epsilon_i^2 + \sum_{i \neq j} c_{ij} \epsilon_i \epsilon_j + \sum_i \frac{\omega_i^2}{2} Q_i^2 + \sum_{k, i \neq j} g_{kij} Q_k Q_i Q_j \\ + \sum_{k,i,j} q_{ijk} \epsilon_i Q_j Q_k \dots, \quad \text{(S11.1)}$$

with $\epsilon_i$ and $Q_i$ representing the amplitudes of strain and phonon eigenmodes, respectively. Note that we adopt the Voigt notation for the strain and hence $\epsilon_1 = \epsilon_{xx}$, $\epsilon_2 = \epsilon_{yy}$, $\epsilon_3 = \epsilon_{zz}$, $\epsilon_4 = 2\epsilon_{yz}$, $\epsilon_5 = 2\epsilon_{xz}$, and $\epsilon_6 = 2\epsilon_{xy}$.

The first two terms in Eq. (S11.1) account for the elastic energy introduced by the rigid deformation of the crystal lattice. The diagonal tensor elements ($c_{ii}$) describe the energetic increase due to the deformation, whereas the off-diagonal elements ($c_{ij}$) contain the Poisson ratio factors, which in general preserve the volume close to its equilibrium value. The following two terms in Eq. (S11.1), containing only the phonon amplitude $Q_i$, account for the energy corresponding to the atomic arrangement within the crystal lattice. Here, quadratic and cubic terms describe harmonic and anharmonic contributions, respectively. In general, the $Q_i$ summation incorporates all phonon branches in reciprocal space. The last term in Eq. (S11.1) is a cross coupling and includes both strain and phonon amplitudes. It mediates changes between the lattice and the underlying atomic arrangement. The specific form of this coupling represents the lowest order term of this type, which becomes non-zero only for certain combinations of strain and phonon eigenvectors. We will later discuss the selection rules for the strain created by this term in detail.

In general, the interplay of the individual terms in Eq. (S11.1) determine the structural phase of a material. Within the harmonic theory of crystals, a structural phase becomes unstable if the phonon eigenfrequencies $\omega_i$ softens. In this scope, strain can also enforce a softening. We find for a single mode, by rearranging the terms in Eq. (S11.1),

$$E_{harm} = \frac{\omega_{eff,i}^2(\epsilon_k)}{2} Q_i^2, \quad \text{(S11.2)}$$

where the effective frequency is

$$\omega_{eff,i}^2(\epsilon_k) = \omega_i^2 + q_{iik}\epsilon_k. \tag{S11.3}$$

The coupling coefficient $q_{iik}$ determines the amount of phonon softening by strain. A large structural susceptibility to strain is often found in perovskite materials, suggesting that they exhibit sizable coefficients $q_{iik}$. SrTiO$_3$ is exemplary in this regard [2–4], as it displays proximity to various structural instabilities such as ferroelectricity or antiferrodistortive states under compressive or tensile strain.

Besides the softening of phonon branches by strain, we can also consider the inverse effect of phonons on the crystal lattice. Noticing that stress is

$$\sigma_i = \frac{\partial E}{\partial \epsilon_i} = c_{ii}\epsilon_i + \sum_{i \neq j} c_{ij}\epsilon_j + \sum_{k,j} q_{ijk}\, Q_j Q_k, \tag{S11.4}$$

where the last term contains explicitly phonon contributions. A single phonon mode hence provides a strain component of

$$\sigma_i^{phonon} = q_{ijj} Q_{jj}^2. \tag{S11.5}$$

By using the elastic compliance tensor, we convert this stress to a strain

$$\epsilon_i^{phonon} = \sum_{j,k,l} s_{ijkl} q_{klj} Q_{jj}^2. \tag{S11.6}$$

Within equilibrium conditions, the thermal occupation of phonon modes given by the Bose-Einstein-distribution governs this contribution to the strain by

$$Q_{jj}^2 = \frac{\hbar}{\omega_j}\left(\frac{1}{e^{\hbar\omega_{ii}/k_B T}-1}+\frac{1}{2}\right). \tag{S11.7}$$

Moreover, if phonon modes become further activated by external stimuli, additional strain components emerge. These extra contributions also act on the structure and can thereby enforce modulations of the crystal lattice. Again, the strain-phonon coupling constants $q_{ijj}$ determine the size of this effect.

To estimate and compute the phonon-induced strain we perform calculations in the framework of density functional theory (DFT). We study the specific case of SrTiO$_3$ in its low-temperature tetragonal phase exhibiting space group *I4/mcm* (140). The external stimuli to displace phonons are light pulses, which only act on phonon modes at the zone center in reciprocal space, $q$ = (0,0,0), because of the small momentum of photons. Moreover, creating sizable displacements of phonons requires matching of photon and phonon frequencies. Consequently, to follow the experimental conditions, we perform our analysis for the highest frequency polar mode in SrTiO$_3$.

To map out all coefficients present in Eq. (S11.1), we have to combine different approaches. Firstly, we deduce the elastic constants utilizing the approach of Ref. [5]. Secondly, the phonon eigenfrequencies and eigenvectors are determined from computations of the force constant matrices for a finite set of symmetry adapted distortions. Finally, we deduce the phonon-strain coupling by superimposing strain and phonon eigenvectors onto the structure of SrTiO$_3$ and fitting the resulting DFT total energy landscape to Eq. (S11.1). A similar approach has also been employed in our previous work (Ref. [6]) and by Vanderbilt *et al.* in Ref. [1].

For our computations, we utilize the implementation of DFT employing the linearized augmented-plane wave method (LAPW) within the ELK-code [7]. The numerical parameters of our calculations have been carefully tested and are the same as in our previous work of Ref. [8]. As an approximation for the exchange-correlation functional we use the generalized gradient approximation in the parametrization of the PBE for solids (PBEsol) functional [9]. All computations are based on a force- and strain-free state which we obtain from structural relaxation of the of SrTiO$_3$ unit-cell. In agreement with other theoretical works [4,8], we find the lattice constants to be $a = b = 5.49$ Å and $c = 7.82$ Å, with the Sr, Ti, O1 and O2 atoms occupying the Wyckoff positions of *b*, c, a and *h*(*x*=0.725), respectively.

For our relaxed SrTiO$_3$ structure, we compute the elastic coefficients by symmetry-adapted least-squares extraction from DFT total energies. In total, there are six independent coefficients for the low temperature structure of SrTiO$_3$, and we list our findings in Table (S11.1), which are in agreement with other database values from DFT [10].

**Table S11.1 | Computed elastic tensor elements for SrTiO$_3$.** Note that we utilize here the Voigt notation.

| C$_{xx}$ | (*GPa*) | C$_{xx}$ | (*GPa*) | C$_{xx}$ | (*GPa*) |
|---|---|---|---|---|---|
| C$_{11}$ | 318 | C$_{12}$ | 93 | C$_{44}$ | 121 |
| C$_{33}$ | 343 | C$_{13}$ | 119 | C$_{12}$ | 74 |

Next, we compute the phonon eigensystem. In total, tetragonal SrTiO$_3$ exhibits 27 non-translational phonon modes at the zone center for which the polar modes span the irreducible representation of $5E_u + 3A_{2u}$, with the $A_{2u}$ modes exhibiting a polarization direction along the tetragonal axis. In Table (S11.2) we show our computed eigen-frequencies, which are in agreement with literature values [4,8]. In addition to the frequencies, we compute the mode effective charge, $Z^*$, for each polar mode to determine the light-phonon mode coupling. We therefore modulate the structure by each phonon-mode eigenvector and calculate the produced ionic and electric polarization. For the latter, we utilize the Wannier90 package [11] to determine the shift of Wannier-Centers accounting for the electronic contribution. Table (S11.2) contains the computed values.

**Table S11.2 | List of the eigenmode frequencies at $q = (0, 0, 0)$ of SrTiO$_3$ computed by DFT.** The assignment of the symmetry of each mode is according to point group *4/mmm*. For the IR-active modes we further list the mode effective charge.

| mode | $f$ (THz) | mode | $f$ (THz) | mode | $f$ (THz) | $Z^*$ (e) |
|---|---|---|---|---|---|---|
| $A_{1g}$ | 3.8 | $B_{2g}$ | 4.3 | $E_u$ | 1.6 $i$ | 3.5 |
| $A_{2g}$ | 14.0 | $B_{2g}$ | 12.8 | $E_u$ | 4.8 | 0.4 |
| $A_{2g}$ | 24.6 | $E_g$ | 1.0 | $E_u$ | 7.3 | 0.1 |
| $A_{1u}$ | 12.7 | $E_g$ | 4.1 | $E_u$ | 12.8 | 0.3 |
| $B_{1g}$ | 14.3 | $E_g$ | 12.9 | $E_u$ | 15.7 | 1.6 |
| $B_{1u}$ | 7.6 | | | $A_{2u}$ | 1.5 $i$ | 3.6 |
| | | | | $A_{2u}$ | 5.0 | 0.0 |
| | | | | $A_{2u}$ | 15.8 | 1.6 |

Last, we compute the coupling term between phonons and strain. Specifically, we consider the contribution of a single polar phonon-mode as written in Eq. (S11.6). The linear-square form of the coupling restricts the possible strain eigenvectors which couple to phonons. Group symmetry considerations dictate that only strain eigenvectors with an irreducible representation of $A_{1g}$ exhibit non-zero coupling coefficients in this case. Consequently, only strains that maintain the symmetry have to be considered. For the tetragonal distorted SrTiO$_3$, there are only two eigenvectors fulfilling this condition. These modes are either a modulation of the in-plane lattice constants by

$$\epsilon_1 = \begin{pmatrix} \chi_1 & 0 & 0 \\ 0 & \chi_1 & 0 \\ 0 & 0 & 0 \end{pmatrix} \tag{S11.8}$$

or a modulation of the tetragonal $c$-lattice constant

$$\epsilon_2 = \begin{pmatrix} 0 & 0 & 0 \\ 0 & 0 & 0 \\ 0 & 0 & \chi_2 \end{pmatrix}. \tag{S11.9}$$

Please note that each strain eigenvector exhibits its own coupling coefficient, but in accordance with Eq. (S11.6), the induced strain scales quadratically with $Q$.

To follow the experimental configuration, we perform calculations for the polar $A_{2u}$ mode at 15.8 THz in combination with both strain vectors. In Fig. (S11.1A), we show the resulting dependence between phonon amplitude $Q$ and induced strain $\epsilon_i$. Our computations show that the $A_{2u}$ distortion creates a tensile stain along the $c$-axis and perpendicular axis, with the former showing the strongest increase. Note, that the strain is tensile for positive and negative amplitudes of $Q_{IR}$. Quantitatively, we obtain as a coupling coefficient 0.02 %/($u\text{Å}^2$) and 0.57 %/($u\text{Å}^2$) for $\epsilon_1$ and $\epsilon_2$, respectively. Please note that $u$ is the atomic mass unit; these coefficients can also be converted into pressure equivalents, which are $0.75\ GPa/(u\text{Å}^2)$ and $2.02\ GPa/(u\text{Å}^2)$, respectively.

To give an estimate of how much strain light pulses produce, we next compute phonon amplitudes. We therefore consider optical pulses with frequencies of 20 THz, tuned close to the optical resonance of the $A_{2u}$ mode, with full width half maximum (FWHM) of 0.15 ps and a field strength between 1 and 18 MV/cm. Our computational model, based on a driven harmonic oscillator, incorporates the mode effective charge of Table (S11.2) and is described in detail in Refs. [6,12]. We find that for these pulse parameters the maximum phonon amplitude is

$$Q_{IR}^{max} = aE, \tag{S11.10}$$

with $a = 0.03$ $(\sqrt{u}\text{Å})/(\text{MV/cm})$. Hence, a driving strength of 18 MV/cm induces a maximum amplitude of $Q_{IR} = 0.5$ $(\sqrt{u}\text{Å})$. Structurally, this amplitude corresponds to a 5 pm displacement of oxygen atoms, which is about 3% of the equilibrium Ti-O bond length.

Last, we consider the full energetic landscape given in Eq. (S11.1) with respect to the phonon induced strain. The aim is to identify the full transient strain state including all contributions. We therefore minimize the total energy with respect to the strain variables for fixed values of phonon induced strain, according to Fig. (S11.1A). Fig. (S11.1B) shows our findings, where we plot the transient strain state as a function of electric field of the mid-infrared light pulses. We find that the transient strain state expands along the tetragonal axis and compresses the in-plane $a$-axis. The latter originates from the volume conservation and Poisson's ratio, which in contrast to the pure phonon induced strain, prefers a state of constant volume. Consequently, to compensate the large strain along the tetragonal axis the system reduces in-plane dimensions. Quantitively, we find that the optical pulses induce a strain along the $c$-axis of up to 0.2%.

A more detailed study of this effect will be published elsewhere.

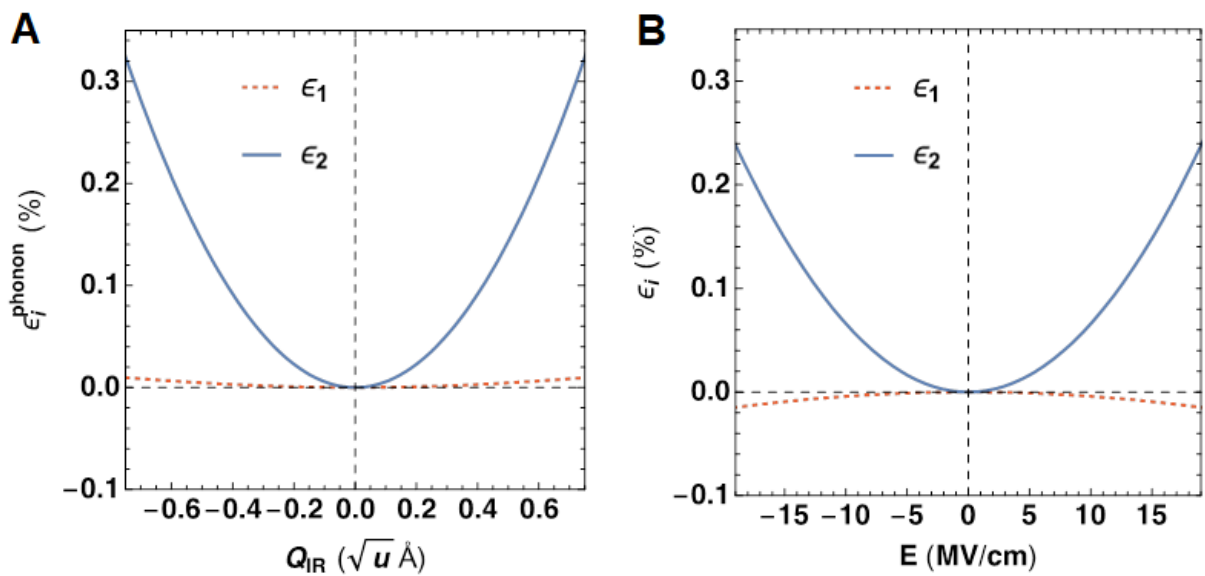

**Figure S11.2 | Computed induced strain as a function of phonon amplitude.** In (A) we show only the phonon induced part; in (B) we show the phonon-induced strain from the full minimization of the free energy in Eq. (S11.1).